\documentclass[epj]{svjour}
\usepackage{graphics}
\begin{document}
\title{Runaway Events Dominate the Heavy Tail of Citation Distributions}

\author{Michael Golosovsky and Sorin Solomon 
\thanks{\emph{e-mail address:} golos@cc.huji.ac.il}%
}                     
\offprints{}          
\institute{The Racah Institute of Physics, The Hebrew University of Jerusalem, Jerusalem 91904, Israel}
\date{Received: date / Revised version: date}
%
\abstract{Statistical distributions with heavy tails are ubiquitous in natural and  social phenomena. Since the entries in heavy tail have unproportional significance, the knowledge of its exact shape  is very important.  Citations of scientific papers form one of the best-known heavy tail distributions. Even in this case there is a considerable debate whether citation distribution  follows the log-normal or power-law fit. The goal of our study is to solve this debate by measuring citation distribution for a very large and homogeneous data. We  measured citation distribution for 418,438 Physics papers published  in 1980-1989 and cited by 2008. While the log-normal fit deviates too strong from the data, the discrete power-law function with the exponent $\gamma=3.15$ does better and fits  $99.955\%$ of the data. However, the extreme tail of the distribution  deviates upward even from the power-law fit and exhibits a dramatic "runaway" behavior. The onset of the runaway regime is  revealed macroscopically as the paper garners 1000-1500  citations, however the microscopic measurements of autocorrelation in citation rates are able to predict this behavior in advance.
} 
\authorrunning{M. Golosovsky and S. Solomon}
\titlerunning{Citation statistics}
\maketitle
\section{Introduction}
\label{intro}
Dynamic statistical distributions found in nature  frequently exhibit heavy tails that are usually approximated using log-normal or power-law functions \cite{Newman}. However, there can be a few individual values in the extreme end of the distribution that exceed by far both these  fitting functions. We advance the hypothesis that these "runaways"  can eventually capture the entire system. Statistical distribution of citations of scientific papers offers a good example of such dynamics.

Intensive studies of  citation statistics were triggered by the  seminal work of de Solla Price \cite{Solla65} who showed that the  statistical distribution of citations of scientific papers has heavy tail  that can be approximated by a power-law function.  De Solla Price proposed a microscopic "cumulative advantage" mechanism \cite{Solla76} that in the long time limit generates such power-law distribution. Following his line of thought, subsequent studies of citation statistics \cite{Redner1998,Lehmann,Borner,Raan2005,Newman2009,Peterson}  used power-law probability distributions to fit their data. Later on, it turned out that  stretched exponential \cite{Wallace,Phillips,Sornette} or log-normal \cite{Redner2005,Amaral,Fortunato,Petersen} functions fit the measured citation distributions equally well.  Ref. \cite{Clauset} showed that the choice between the power-law and the log-normal fits  is extremely difficult. Indeed, the most important difference between the two is that the power-law decays slower than the log-normal, in other words it has  heavier tail. To probe this tail one has to measure a very large data set in order to achieve enough entries in the tail. Previous studies of citations of scientific papers used either  small  \cite{Amaral,Fortunato} or inhomogeneous \cite{Redner2005} data sets that were insufficient to distinguish between the log-normal and power-law fits. 

 The initial goal of our  study was to find out which function  fits better the citation distribution in the long  time limit: the power-law or the log-normal. To achieve this goal we measured citation distribution for a very large and homogeneous data set containing almost all  Physics papers published in 1980-1989 and cited by 2008. Surprisingly, we found that (i) the citation distribution is not stationary even 25 years after publication, and (ii) both log-normal and power-law fitting functions fail to describe the extreme tail of the distribution since it develops a runaway behavior.

\section{Theoretical framework}
In the long time limit the de Solla Price's Cumulative Advantage mechanism \cite{Solla76} generates the following citation distribution
\begin{equation}
p(k)=\frac{B(k+w,\gamma)}{B(w,\gamma-1)}.
\label{Yule}
\end{equation}
Here, $k$ is the number of citations, $B(a,b)=\frac{\Gamma(a)\Gamma(b)}{\Gamma(a+b)}$ is the beta-function, $\Gamma$ is the gamma-function,  and $w,\gamma$ are parameters.  The Eq.\ref{Yule} is  known   as Waring \cite{Burrell2005,Mingers,Glanzel}) or  discrete power-law  distribution  since for $k>>w$ it asymptotically approaches the power-law dependence, $p(k)\sim\left(w/k\right)^{\gamma}$. 
The continuous approximation of Eq.\ref{Yule} 
\begin{equation}
p(k)\approx (\gamma -1)\frac{w^{\gamma-1}}{(w+k)^{\gamma}}
\label{Pareto}
\end{equation} 
is known as the  Zipf-Mandelbrot or Pareto-type II distribution  and was also used in citation analysis \cite{Wallace,Glanzel,Tsallis}.

\begin{figure}[ht]
\resizebox{0.45\textwidth}{!}
{\includegraphics*{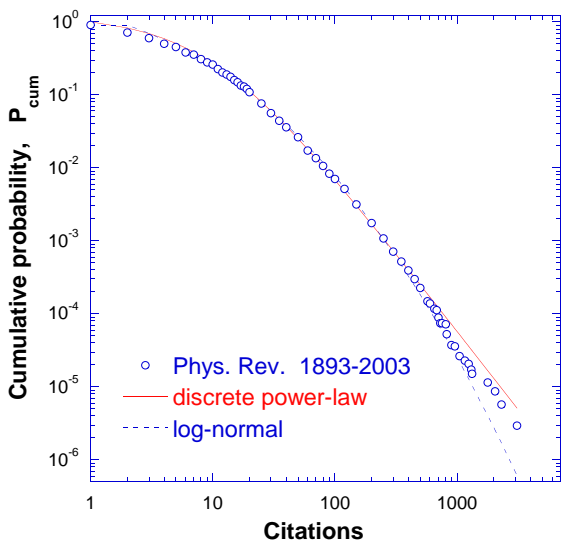}}
\caption{Cumulative probability distribution (cdf) of citations  to 353,268 papers published in Physical Review journals during 1893-2003 and cited  by 2003. Only PR to PR citations were counted. The data were adapted from Ref. \cite{Redner2005}.  The continuous red line shows a fit with the discrete-power-law cdf (Eq.\ref{Yule}) with $\gamma=3.15,w=10.2$. The dashed blue line shows a fit with the  log-normal cdf (Eq.\ref{lognormal}) with $\mu=1.15,\sigma=1.42$.
}
\label{fig:redner}
\end{figure}

Several recent works \cite{Redner2005,Amaral,Fortunato,Petersen} fitted measured citation distribution with the log-normal 
\begin{equation}
p(k)=\frac{1}{k\sigma\sqrt{2\pi}} e^{-\frac{(\ln k-\mu)^{2}}{2\sigma^{2}}}
\label{lognormal}
\end{equation}
function that was modified to describe such non-negative and discrete variable as citations. Although the functions represented by Eqs.\ref{Yule},\ref{lognormal} are dramatically different, the fits of the citation distributions based on Eq.\ref{Yule} and on the discrete version of Eq.\ref{lognormal} are  virtually undistinguishable  for a wide (but finite!) range of values. Thus, discrimination between the discrete log-normal and the discrete power-law distributions is quite ambiguous \cite{Clauset,Egghe}. This ambiguity is illustrated by Fig. \ref{fig:redner} that shows cumulative probability distribution of citations  to 353,268 Physics papers. The  data lie just in between  the log-normal cdf and discrete power-law cdf.

The solution of the dilemma "power-law or log-normal" comes from an unexpected direction. Our measurements demonstrate that (i) the  citation distribution  follows neither power-law nor log-normal fit, (ii) it is non-stationary, and (iii) the individual values in the tail  grow at a much faster rate than the rest of the distribution, indicating the runaway effect. As the sample size and the evolution time increase, the runaways  become more prominent. The contribution of non-stationarity in the emergence of distributions with heavy and super-heavy tails  has been studied in the context of  cities population, wealth distribution and growing networks \cite{Blank,Malcai,Klass}.

\section{Citation Distribution  for Physics Papers}
We considered research papers in Physics  that were published in 82 leading Physical journals in the period from 1980 to 1989 (excluding the overview papers and papers published in the popular science journals) -418,438 papers in total. We measured the  number of citations gained by each  paper by July 2008.  Since the citation life of an ordinary paper rarely exceeds 15 years and the time span between the publication and  citation count for the papers in our data set is 20-28 years, we expected  that the citation distribution for this data set of "old" papers is stationary.

\begin{figure}[ht]
\resizebox{0.5\textwidth}{!}
{\includegraphics*{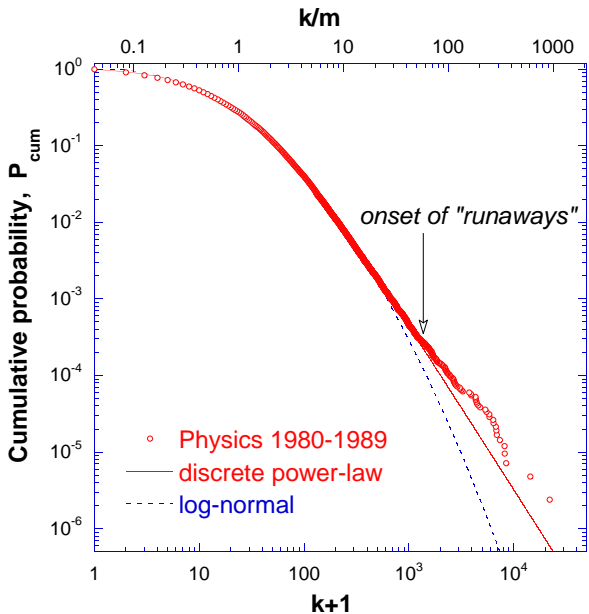}}
\caption{Cumulative  distribution function (cdf) of citations to 418,438  papers published in 82 leading Physical journals in 1980-1989  and cited by July 2008. The horizontal axis is $k+1$ instead of $k$ in order to show uncited papers, $k=0$, on the log-scale. The dashed blue line shows a fit with the log-normal cdf (Eq.\ref{lognormal}) with $\mu=2.16$ and $\sigma=1.38$. The deviation from the fit appears already at $k>400$. The continuous red line shows a  fit with the  discrete-power-law cdf (Eq. \ref{Yule}) with $\gamma=3.15$ and $w=27.5$. This function fits the data up to $k=1000$. However,  the tail of the distribution with $k>1000$ deviates upwards even from this power-law fit, indicating "runaway" papers.
}
\label{fig:all}
\end{figure}

\begin{figure}[ht]
\resizebox{0.5\textwidth}{!}
{\includegraphics*{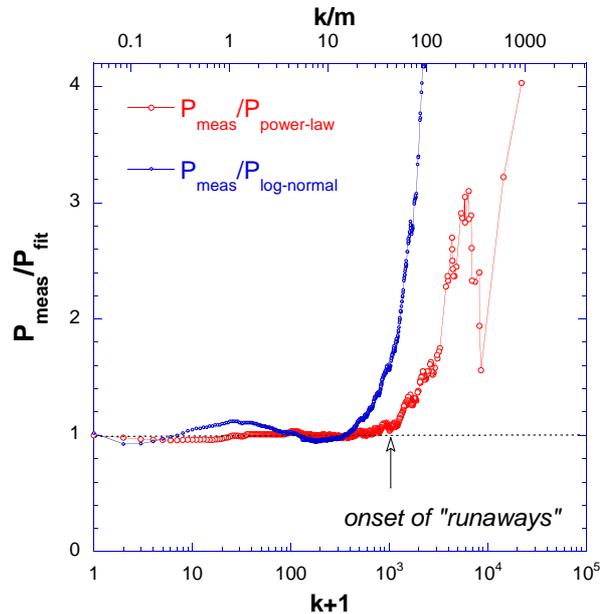}}
\caption{The ratio of the measured cumulative probability of citations (Fig.\ref{fig:all}) to the discrete power-law cdf (red circles) and to the log-normal cdf (blue circles). The log-normal fit fails already for $k>400$. The discrete power-law provides a good fit for $k<1000$ while pronounced deviation appears above $k>1000$. 
}
\label{fig:ratio}
\end{figure}

Figure \ref{fig:all} displays this  distribution. The log-normal cdf  fits the data only for the papers with less than 400 citations  (these constitute 99.7$\%$ of all papers), while the tail  deviates upward. 
The  discrete power-law  cdf  fits the data in Fig. \ref{fig:all}  much better. However, it also fails to describe the extreme tail of the distribution that contains the papers with more than 1000 citations (0.04 $\%$ of all papers). This  is most dramatically illustrated in Fig.\ref{fig:ratio} where we plotted the ratio of the measured number of citations  to the fitted values using log-normal cdf and discrete power-law cdf. The number of citations for  \emph{all} the 190 papers that have more than 1000 citations significantly and increasingly exceed both the power-law and the log-normal fits.   It is instructive to compare the onset of this super-heavy or "runaway" tail  to the mean number of citations, $m=24$. Figures \ref{fig:all},\ref{fig:ratio} indicate that the  "runaways" become visible at $k/m= 20-40$ and extend at least to $k/m= 1000$. The previous smaller scale studies  \cite{Amaral,Fortunato} were limited to $k/m<40$ and didn't provide enough statistics to probe the "runaways".

We conclude that citation distribution for the small homogeneous data sets ($\sim$1000 papers) can be fitted equally well by the log-normal and by the discrete power-law functions. Larger sets ($\sim$10,000 papers) are better fitted by the discrete power-law cdf. The runaway tail that exceeds even the power-law fit  becomes apparent when the number of papers in the set exceeds 10,000. In what follows we focus on the microscopic mechanism responsible for runaways.

\section{Microscopic mechanisms of  citation dynamics that are able to produce the runaway behavior}
\begin{enumerate}
    \item \emph{Superlinear Yule-Simon process}. The most accepted microscopic mechanism of citation dynamics  is the preferential attachment  also known as the cumulative advantage, Gibrat, or Yule process \cite{Solla76,Barabasi,Krapivsky2000,Dorogovtsev2000}. This mechanism assumes that the citation process is a memoryless Markov chain whereas  the citation rate of a paper is determined by the total number of accumulated citations $k$ and the time after publication $t$,
\begin{equation}
\frac{dk}{dt}=A(t)(k+k_0)^{\alpha}\label{PA}
\end{equation}
Here,  $A$ is the aging parameter, $\alpha$ is the attachment exponent, and $k_0$ is the  initial attractiveness. The Refs. \cite{Krapivsky2000,Dorogovtsev2000} showed that  the linear preferential attachment, $\alpha=1$ yields the discrete power-law citation distribution (Eq.\ref{Yule}) while the superlinear preferential attachment, $\alpha>1$, results in  the runaway behavior where a few papers contain a finite fraction of all citations. 

\item \emph{Stochasticity.} The heavy tail distribution can result from  the stochastic and non-stationary character of the dynamic variable \cite{Blank,Malcai,Levy}. For  dynamic variable  with lower bound, such as citations, the stochastic mechanism yields the power-law distribution whose exponent is the ratio between the average growth rate and its individual fluctuations \cite{Malcai,Klass}. The finite size effects may affect the power-law exponent to the extent of concentrating the entire distribution in just a few individual entries in the end of the heavy-tail \cite{Blank,Huang}.  Hence, the stochasticity can partially explain the power-law citation distribution and runaway behavior. 

\item \emph{Memory.}  If the assumption of a memoryless Markov chain that stands behind Eq.\ref{PA} is lifted (in other words, citation process has some memory) this can  result in the distribution with the runaway tail. Even a weak memory of the past dynamics and its feedback on the rate of the multiplicative random walk may lead to divergent statistical distribution and runaways \cite{Louzoun}. 
\end{enumerate}  
   
To verify to which these mechanisms shape the citation distribution we measured  citation dynamics of individual papers. We expected that once the microscopic dynamics is uncovered, we will be able to explain the "runaway tail".

\subsection{Microscopic measurements of citation dynamics of individual papers} 
In distinction to the measurements shown in Fig.\ref{fig:all} where we measured aggregated citation distribution for the 418, 438 Physics papers published during a decade (1980-1989) and didn't trace citation dynamics of individual papers; here we considered the Physics papers published in \emph{one} year (1984) and analyzed citation history of \emph{each} paper up to July 2008. This data set contains a smaller number of entries (40,195) but it is  age-homogeneous.

\subsubsection{Divergence of citation dynamics of similar papers}
\begin{figure}[ht]
\resizebox{0.45\textwidth}{!}
{\includegraphics*{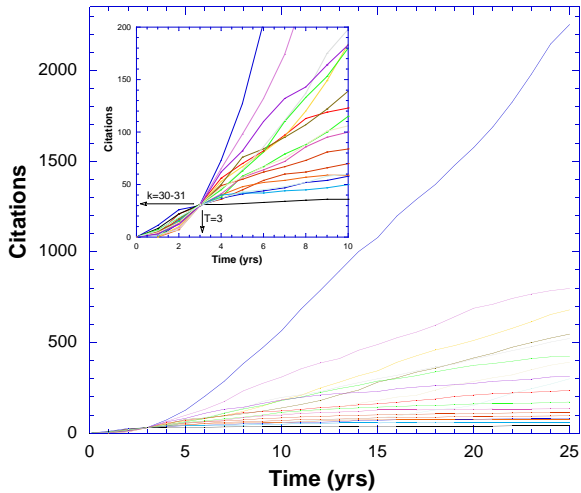}}
\caption{Citation dynamics of 89 Physics papers published in 1984. We chose all those papers that by 1986 (three years after publication) had 30 or 31 citations. Although the initial citation dynamics of these papers is very similar, it quickly diverges in such a way that after 25 years (in 2008) the number of citations varies between 40 and 2254. 
}
\label{fig:trajectory}
\end{figure}

 As an example, from all Physics papers published in 1984 we  chose the subset  containing all those papers that garnered 30-31 citations by the end of 1986. Figure \ref{fig:trajectory} displays citation dynamics of these 89 papers. Although they represent an extremely homogeneous subset (the same field, the same publication year, the same citation prehistory), the divergence in their citation behavior is striking.

We ranked the papers according to the number of citations garnered by 2008. The list opens with the pair of papers:  "Lower critical dimension of the random-field Ising model- a Monte-Carlo study" by  D. Andelman, H. Orland, and L.C.R. Wijwardhana, Phys.Rev. Lett.  (40 citations); and "Fractional  quantum Hall-effect at filling factors up to $\nu= 3$" by G.  Ebert, K. von Klitzing, and J.C.Maan, J. of Physics C- Solid State  (49 citations). The lists ends with the papers: "Dynamics of supercooled liquids and the glass transition" by U. Bengtzelius, W. Gotze and A. Sjolander, J. of Physics C- Solid State (798 citations); and "Embedded-atom method: Derivation and application to impurities, surfaces, and other defects in metals" by M.S.  Daw and M.I. Baskes, Phys. Rev. B (2254  citations). All four papers were published actually in the same journals in the same year and they have the same citation prehistory. The two former papers are important works with typical citation dynamics. The two latter papers exhibit strongly different  citation dynamics  and the last one  is a clear runaway. S.Redner \cite{Redner2005} already demonstrated that the citation history of the citations classics is strongly individualized. Our  analysis substantiates this observation and extends it to all papers.

\subsubsection{Analysis of the citation dynamics}
We analyzed citation dynamics of all 40,195 Physics papers published in 1984 using the framework of Eq.\ref{PA}. We found $\alpha=1.2-1.28, k_{0}=1.1$ and $A=3.5/(t+0.3)^{2}$.  Since the deviation from the linearity is small, $\alpha-1<<1$,  and the aging parameter $A$ strongly decays with time, the runaway behavior generated by Eq.\ref{PA} is too slow.  We conclude that while the nonlinear preferential attachment is the dominant mechanism that shapes  the observed power-law citation distribution, it can not produce runaways.  

As is clearly seen from the Fig.\ref{fig:trajectory}, the fluctuations of the citation rate of individual papers (the ripple on the continuous lines) is small.  Therefore, while the stochasticity affects the shape of the citation distribution  its contribution is not dominant. Neither it can be responsible for runaways.

The strong systematic differences between the citation behavior of similar papers shown in Fig.\ref{fig:trajectory} is at odds with the preferential attachment paradigm  that assumes  similar  citation dynamics for the papers having the same past number of citations. As borne out by the data, there are strong correlations between the citation rates of the same paper in different years, i.e. the citation process has memory \cite{Cattuto}. As we  show below, the memory dominates the dynamics of heavily cited papers and turns the citation process in a \emph{de facto} deterministic one, at least for the runaway papers.   Indeed, Fig. \ref{fig:trajectory}  shows that  the number of citations for the most part of the papers comes to saturation after 15 years. However, some papers continue to be cited with undiminished rate even  after 25 years.  It is clearly seen that the citation behavior of these runaway papers is not erratic, the  citation rates in subsequent years are strongly correlated.

\subsubsection{Autocorrelation}
To characterize the correlations we chose the sets of similar papers and measured autocorrelation between the citation rates of subsequent years. To this end we considered the sets of papers that were published in the same year and that garnered the same number of citations, $k$, after $t$ years. For each  paper $i$ we found the number of citations acquired during two subsequent years - $\Delta k_{i}(t)$ and $\Delta k_{i}(t-1)$- and calculated the Pearson autocorrelation coefficient, 
\begin{equation}
c_{t,t-1}=\frac{\overline{\left(\Delta k_{i}(t)-\overline{\Delta k_{i}(t)}\right)\left(\Delta k_{i}(t-1)-\overline{\Delta k_{i}(t-1)}\right)}}{\sigma_{t}\sigma_{t-1}}
\label{c}
\end{equation}
Here  $\sigma_{t},\sigma_{t-1}$ are standard deviations of the $\Delta k_{i}(t)$ and $\Delta k_{i}(t-1)$ distributions, respectively; and  the averaging is performed over all papers in the set.  
\begin{figure}[ht]
\resizebox{0.5\textwidth}{!}
{\includegraphics*{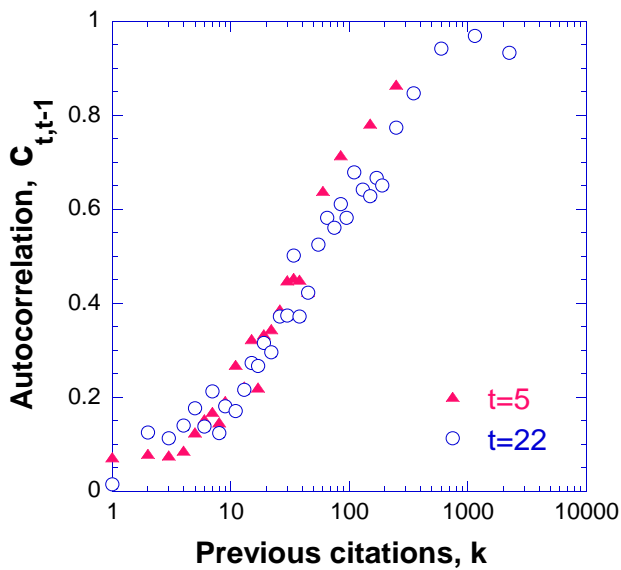}}
\caption{ The Pearson autocorrelation coefficient  for additional citations (Eq.\ref{c}). $t$ is the number of years after publication. $c_{t,t-1}$ steadily increases with the number of previous citations $k$, and exceeds 0.9 for the papers with more than 1000 citations.
}
\label{fig:correlation}
\end{figure}
Figure \ref{fig:correlation} shows that for moderately cited papers the autocorrelation is weak, $c_{t,t-1}<<1$, as expected for a memoryless process. However, the papers with more than 1000 citations have $c>0.9$. This means that their citation behavior is almost deterministic. While the citation rate of the most part of the papers decays with time, these papers continue to be highly-cited and eventually  develop into runaways.  (Indeed, the onset of the runaway tail in Fig.\ref{fig:ratio} occurs at the same number of citations, $k=1000$,  at which the autocorrelation coefficient approaches unity.) In the long run, the citation rate of these runaways shall nevertheless decrease due to yet another mechanism - these prominent papers eventually become common knowledge and need not to be cited anymore.

\subsubsection{Numerical simulations}
Our further studies  of citation statistics (not shown here) combined correlation measurements with  other parameters of the correlated Poisson process governing the citation dynamics of each paper, namely, stochastic and deterministic parts of the Eq.\ref{PA}. We performed numerical simulations of citation dynamics based on these measurements and without any additional parameters, and found aggregated citation distribution.  The simulations developed a runaway population of papers consistent with that observed in Fig.\ref{fig:all}.  The simulations based only on the stochastic version of the Eq.\ref{PA} (without correlations) yielded discrete power-law citation distribution and no runaways. 

\section{Discussion}
Unlike other fields where "dragon king" events have negative emotional connotation and are associated with catastrophes and crashes, the  runaways  in citation statistics have positive connotation and indicate very important papers in the field.  The social and epistemological implications of our study might contribute to the current effort of evaluating and predicting scientific production in terms of citations.

Our results are important also in a broader context of the dynamics of complex systems. Our citation statistics results add themselves to the relatively small number of systems for which the microscopic elementary laws were uncovered and whose macroscopic consequences (obtained theoretically and by simulation) passed successfully the confrontation with the empirically observed macroscopic phenomena \cite{Gur,Challet,Dover}.

The present study should be viewed as a part of the more generic effort of understanding autocatalytic dynamics as the bridge that allows the promotion of simple microscopic interactions to complex, macroscopic collective phenomena.  

\section{Conclusions}

\begin{itemize}

\item[1] Aggregate citation distribution for the set of "old" Physics papers is better fitted by the discrete power-law (Waring) distribution rather than by the log-normal distribution. \\

\item[2] Both power-law and log-normal fits severely underestimate the extreme tail of the citation distribution that contain papers with more than few thousands citations ($\sim 0.05\%$ of all papers). These "runaways"  exhibit significant autocorrelation in their citation rate (the Pearson autocorrelation coefficient exceeds 0.9.) in such a way that it becomes deterministic.\\

\item[3]In the long run, citation distribution for the age-and field-homogeneous set of papers reveals two populations: ordinary papers whose citations come to saturation after 15-20 years and  "runaways" that continue to be cited even after 20 years.\\

\end{itemize}

\section{Acknowledgments}
We are grateful to B. Meerson, G.J. Peterson, S. Redner, P. Richmond, N. Shnerb, and J. Weissbuch   for valuable discussions.  We appreciate important   correspondence with Q. Burrell, S. Fortunato, and J.C. Phillips.  This work was partly supported by the CO3, Daphnet.

%

\end{document}